\documentclass[showpacs,amsmath,amssymb,aps,showkeys,floatfix,a4paper]{revtex4}

\usepackage{dcolumn}
\usepackage{bm}
\usepackage{epsfig}
\usepackage{amsfonts}
\usepackage{amssymb,amscd}
\usepackage{graphicx}
\usepackage{wrapfig}

\def\lsim{\raise0.3ex\hbox{$<$\kern-0.75em\raise-1.1ex\hbox{$\sim$}}}
\def\gsim{\raise0.3ex\hbox{$>$\kern-0.75em\raise-1.1ex\hbox{$\sim$}}}

\def\beq{\begin{equation}}
\def\eeq{\end{equation}}
\def\bea{\begin{eqnarray}}
\def\eea{\end{eqnarray}}
\def\bq{\begin{quote}}
\def\eq{\end{quote}}
\newcommand{\sigmatot}{\sigma_\text{tot}}

\def\gappeq{\mathrel{\rlap {\raise.5ex\hbox{$>$}}
{\lower.5ex\hbox{$\sim$}}}}

\def\lappeq{\mathrel{\rlap{\raise.5ex\hbox{$<$}}
{\lower.5ex\hbox{$\sim$}}}}

\def\Toprel#1\over#2{\mathrel{\mathop{#2}\limits^{#1}}}

\begin{document}


\title{The Phillips - Barger model for the elastic cross section and the Odderon}
\author{V.~P. Gon\c{c}alves}
\email{barros@ufpel.edu.br}
\affiliation{High and Medium Energy Group, \\
Instituto de F\'{\i}sica e Matem\'atica, Universidade Federal de Pelotas\\
Caixa Postal 354, CEP 96010-900, Pelotas, RS, Brazil}
\author{P. V. R. G. Silva}
\email{precchia@ifi.unicamp.br}
\affiliation{High and Medium Energy Group, \\
Instituto de F\'{\i}sica e Matem\'atica, Universidade Federal de Pelotas\\
Caixa Postal 354, CEP 96010-900, Pelotas, RS, Brazil}
\date{\today}

\begin{abstract}
Inspired by the recent TOTEM data for the elastic proton -- proton ($pp$) scattering at $\sqrt{s} =$ 8 and 13 TeV, we update previous studies of the differential cross sections using the Phillips -- Barger (PB) model, which parametrizes the amplitude in terms of a small number of free parameters. We demonstrate that this model is able to describe the recent $pp$ data on a statistically acceptable way. Additionally, we perform separate fits of the $pp$ data for each center - of - mass energy and propose a parametrization for the energy dependence of the parameters present in the PB model. As a consequence, we are able to present the PB predictions for the elastic proton - proton cross section at $\sqrt{s} = 546$ GeV and $1.8$ TeV, which are compared with the existing antiproton -- proton ($\bar{p}p$)  data. We show that the PB predictions, constrained by the $pp$ data,  are not able to describe the $\bar{p}p$ data. In particular, the PB model predicts a dip in the differential cross section that is not present in the $\bar{p}p$ data. Such result suggests the contribution of the Odderon exchange at high energies.
\end{abstract}


\maketitle


The  LHC measurements for the total, elastic and differential cross sections at high energies of 2.76, 7, 8 and 13 TeV  
\cite{Antchev2011,Antchev2013,Antchev2015,Antchev2016,Deile2017,Antchev2017b,Antchev2018a,Antchev2018b,
Antchev2017,Aad:2014,Aaboud:2016} provide very interesting results, which challenge the description  of the theory of strong interactions at high energies. In particular, the high-precision measurement
of the elastic  differential cross section data ($d\sigma/dt$) 
in the Coulomb-Nuclear interference region in proton-proton ($pp$) scattering,  performed by the TOTEM Collaboration \cite{Antchev2017}, 
has allowed the determination of the $\rho$ parameter, which is defined as the
ratio between the real and imaginary part of the elastic scattering amplitude 
in the forward direction. The value obtained experimentally is lower than those predicted by different Regge inspired models that take into account  the Reggeon and Pomeron exchange contributions for the scattering amplitude,
whose predictions for the total cross section ($\sigmatot$) are compatible with the measurements \cite{Pancheri:2016yel}. Such result has motivated an intense debate about the possible contribution of the Odderon \cite{
Martynov:2017zjz, Khoze:2018bus,Broilo:2018els, Troshin:2018ihb,Broniowski:2018xbg,Broilo:2018qqs,Gotsman:2018buo,Csorgo:2018uyp}. 
The existence of an Odderon is a natural prediction of the Quantum Chromodynamics (QCD), has a $C$-odd parity  and determines the hadronic cross section difference between the direct and crossed channel processes at very high energies (for a review see Ref. \cite{Ewerz}). If the Odderon exchange contributes for the hadronic scattering at high energies we will have $\sigmatot^{pp}(\sqrt{s}) \neq \sigmatot^{\bar{p}p}(\sqrt{s})$. Moreover, 
the squared momentum transfer ($t$) dependence of  $d\sigma/dt$ for $|t|\neq 0$ is predicted to be different for $pp$ and $\bar{p}p$ scattering at a fixed center - of - mass energy ($\sqrt{s}$) 
\cite{dosch_ewerz,Ster:2015esa}. Namely, a dip is expected to be present in $pp$ and absent in $\bar{p}p$ collisions, which is directly associated to the fact that the Odderon contribution, being a $C$-odd term in the elastic scattering amplitude,  enters with an opposite sign in each case,
resulting in a different $t$ --  behaviour for  $d\sigma/dt$. Consequently, a comparison between the data for the $pp$ and $\bar{p}p$ distributions {\it at the same energy} can be considered a direct probe of the Odderon. Such comparison have only  been performed using the data at low energies ($\sqrt{s} = 53$ GeV), but due to the low statistics of the $\bar{p}p$ data, the difference between the $pp$ and $\bar{p}p$ data is not very large and depends on only a few data points. Therefore, the interpretation of the difference as a evidence of the Odderon remains a subject of debate \cite{Ewerz}.

Another important aspect of the recent TOTEM data for the transverse momentum distributions is its high precision and the large range of $t$ values spanned at high energies. The description  of the $t$ - dependence of $d\sigma/dt$  in the full kinematical range is still a theoretical challenge, since the model should be able to describe the optical point, the forward peak, the dip and the tail for a fixed energy, as well as to predict the change of these structures with the energy. As the elastic process is dominated by non -- perturbative physics, such process have been described using  phenomenological models based on different assumptions. Our goal in this letter is to update the Phillips -- Barger (PB) model, proposed originally in Ref. \cite{Phillips1973}  and modified in Ref. \cite{Fagundes2013}, which describes the scattering amplitude in terms of a small number of free parameters.
As demonstrated in Ref. \cite{Fagundes2013}, this model is able to describe the LHC data at $\sqrt{s} = 7$ TeV. In what follows we extend the analysis for $\sqrt{s} = $ 2.76, 8 and 13 TeV and demonstrate that the PB model also allows to successfully describe the recent TOTEM data. A shortcoming of the PB model is that the energy dependence of the parameters present in this model is not predicted. In this letter we will overcome this limitation using the $pp$ data   
for ISR ~\cite{AmaldiSchubert1980,Schubert1979}
and LHC  \cite{Antchev2011,Antchev2013,Deile2017,Antchev2018a,Antchev2018b} energies to determine the energy dependence of the BP parameters. As a consequence, this procedure allow us to predict  
$d\sigma/dt$ for $pp$ scattering at FERMILAB energies, where $\bar{p}p$ data exist. As we will show below, the PB model, constrained by $pp$ data, is not able to describe the FERMILAB data at $\sqrt{s}$ = 546 GeV and 1.8 TeV. Such result suggests the presence of the Odderon in the description of the hadronic scattering at high energies. 
 
Initially let's present a brief review of the modified version of Phillips -- Barger model proposed in Ref. \cite{Fagundes2013}. In this model the  elastic scattering amplitude is parametrized as follows
\begin{equation}
 \mathcal{A}_\text{PB}(s,t) = i[F_p^2\sqrt{A}e^{Bt/2} + e^{i\phi}\sqrt{C}e^{Dt/2}],
 \label{eq:pb_modified}
\end{equation}
where $F_p^2(t) = 1/(1-t/t_0)^4$ is the proton form factor and $t_0$, 
$\sqrt{A}$, $\sqrt{C}$, $B$, $D$, $\phi$ are free parameters. 
The simple functional form of Eq.~\eqref{eq:pb_modified}
allows to identify the role of each parameter: $t_0$ is important for the description of the data at small values of $|t| \approx 0$ 
(excluding the Coulomb-Nuclear interference region),   
$\sqrt{A}$ and $B$ are relevant in the diffraction peak region,  while 
$\sqrt{C}$ and $D$ are important to describe the data beyond the dip. 
Finally, the phase $\phi$, which enters in the interference 
term appearing in $|\mathcal{A}_\text{PB}|^2$,
controls the position and depth of the dip. 
It is important to emphasize that the PB amplitude cannot be expressed in terms of even and odd contributions, as usually assumed in the phenomenological analysis of the elastic scattering. The two terms in Eq.~\eqref{eq:pb_modified} receive contributions from different charge conjugation processes. In particular, the second term has contributions from both $C = \pm 1$ terms \cite{Fagundes2013}. The separation of the two contributions can only be performed if additional assumptions are included in the analysis, which become the final interpretation strongly model dependent. In our study, we will not consider this possible approach, but instead to assume the PB amplitude as an empirical parametrization that captures the main aspects of the elastic scattering. We will apply this model for $pp$ and $\bar{p}p$ collisions, separately, and compare the final results for a same energy. Our idea is to verify if the resulting predictions are similar or distinct. If distinct, we have an indication that the $pp$ and $\bar{p}p$ amplitudes receive different contributions, with the Odderon contribution being one of the possible sources of this difference. Certaintly, this aspect deserves more detailed studies in the future.

As pointed out before, such model does not predict the energy dependence of the parameters. In order to determine this dependence we will assume the following strategy in  what follows:  we will fit the $d\sigma/dt$ data separately for each energy  using Eq.~\eqref{eq:pb_modified} and the normalization $ {d\sigma}/{dt} = \pi |\mathcal{A}_\text{PB}|^2$. After we will fit the parameters appearing in the modified PB model by empirical functions of $s$. 
We will include in our analysis the $pp$ data at
ISR energies ($\sqrt{s} =$~23, 31, 45, 53 and 63 GeV)~\cite{AmaldiSchubert1980,Schubert1979}
and the LHC data for $\sqrt{s} =$~2.76, 7, 8 and 13~TeV
obtained by the TOTEM Collaboration \cite{Antchev2011,Antchev2013,Antchev2015,Antchev2016,Antchev2017,Antchev2018a,Antchev2018b}.
In order to cover a larger range of values of $t$, 
we also include the preliminary TOTEM data  for 8 TeV \cite{Deile2017}.
For all energies considered, the optical point was included in the dataset, 
so that our fits are constrained by the data for the total cross sections.
Finally, we have constrained the $\phi$ parameter to be in 
the 3rd quadrant in order to obtain
the correct signal of the real part of the amplitude 
and, consequently, to be in accordance with the signal of
the experimental information available for the $\rho$ parameter. 
This constraint does not affect the other parameters 
or the description of $d\sigma/dt$ data, since $\phi$ 
appears inside a cosine function in the interference term of $|\mathcal{A}_\text{PB}|^2$.

\begin{table}[t]
 \centering
  \begin{tabular}{|c||ccccc|}\hline\hline
 \multicolumn{6}{|c|}{$pp$ scattering - ISR energies}\\\hline
Energy             & 23 GeV    & 31 GeV    & 45 GeV     & 53 GeV     & 63 GeV     \\\hline
$A$                 & 24.42(18) & 26.09(10) & 28.528(46) & 29.658(46) & 30.706(68) \\
$B$                 & 3.81(11)  & 4.132(79) & 4.258(54)  & 4.116(49)  & 4.58(11)   \\
$C (\times 10^{-3})$& 1.80(26)  & 1.48(14)  & 0.772(50)  & 1.101(52)  & 0.900(95)  \\
$D$                 & 2.089(55) & 2.070(35) & 1.837(24)  & 1.920(14)  & 1.910(38)  \\
$\phi$              & 3.296(11) & 3.225(12) & 3.449(12)  & 3.446(11)  & 3.459(26)  \\
$t_0$               & 1.016(22) & 1.050(16) & 0.9100(86) & 0.9027(74) & 0.929(15)  \\\hline
$\chi^2/\nu$        & 1.24      & 1.218     & 3.689      & 2.864      & 1.245      \\       
$\nu$               & 128       & 167       & 202        & 200        & 119        \\\hline\hline
\end{tabular}
\caption{ Results for the fits of the  differential 
 cross section data considering $pp$ scattering at ISR energies. The reduced $\chi^2$ and degrees of freedom, $\nu$, for each fit are also presented.}
\label{tab:res-dsigdt-ISR}
 \end{table}

\begin{table}[t]
 \centering
\begin{tabular}{|c||cccc|}\hline\hline
 \multicolumn{5}{|c|}{$pp$ scattering - LHC energies}\\\hline
Energy             & 2.76 TeV     & 7 TeV      & 8 TeV      & 13 TeV      \\\hline
$A$                 & 135.3(3.5)   & 177.42(53) & 201.1(6.2) & 238.64(24)  \\
$B$                 & 6.62(75)     & 8.69(17)   & 5.94(63)   & 7.155(28)   \\
$C (\times 10^{-3})$& 150 (fixed)  & 406(19)    & 1149(456)  & 1484(14)    \\
$D$                 & 3.35(24)     & 4.609(39)  & 5.20(36)   & 5.4751(79)  \\
$\phi$              & 3.571(39)    & 3.5366(76) & 3.393(31)  & 3.41731(91) \\
$t_0$               & 0.709(67)    & 0.727(12)  & 0.602(22)  & 0.6285(14)  \\\hline
$\chi^2/\nu$        & 1.097        & 2.626      & 0.809      & 6.30        \\       
$\nu$               & 59           & 156        & 84         & 398         \\\hline\hline
\end{tabular}
\caption{Results for the fits of the  differential 
 cross section data considering $pp$ scattering  at LHC energies. The reduced $\chi^2$ and degrees of freedom, $\nu$, for each fit are also presented.}
 \label{tab:res-dsigdt-LHC}
 \end{table}

\begin{figure}[t]
 \centering
 \includegraphics[width=0.45\textwidth]{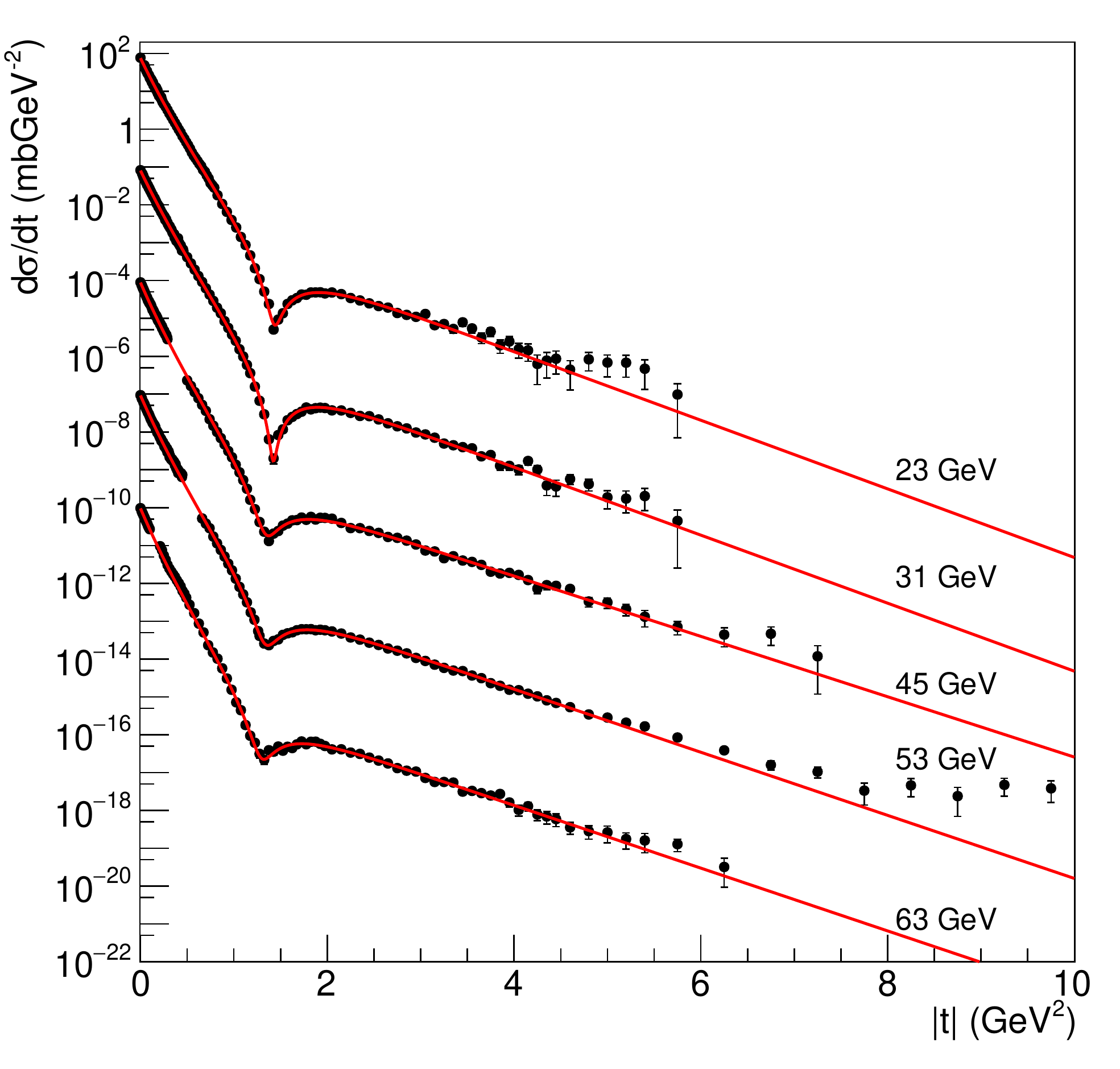}
 \includegraphics[width=0.45\textwidth]{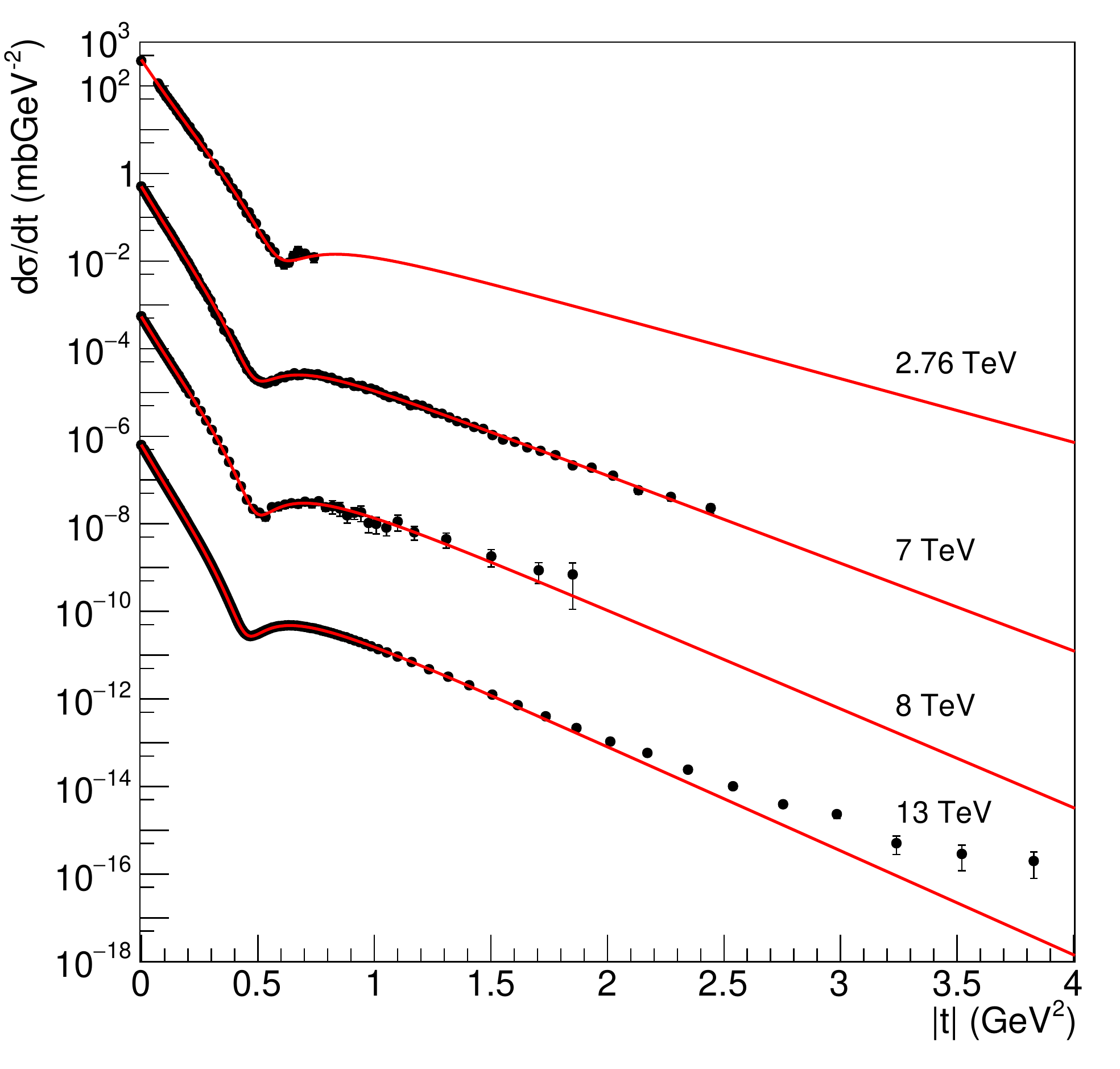}
 \caption{\label{fig:dsigdt-pp} Comparison between the PB results and the experimental data for  $d\sigma/dt$ in  $pp$ collisions at ISR (left) and LHC (right) energies.}
\end{figure}

The fit results are shown in the Tables~\ref{tab:res-dsigdt-ISR} and \ref{tab:res-dsigdt-LHC} for $pp$ collisions at ISR and LHC energies, respectively,  as well as the reduced $\chi^2$ and degrees of freedom, $\nu$, for each fit. Parameters $A$ and $C$ are given in mb$^2$GeV$^{-4}$, $B$ 
 and $D$ in GeV$^{-2}$, $t_0$ in GeV$^2$ and $\phi$ in radians. 
The quoted uncertainties correspond to 1$\sigma$ of confidence level.
The comparison between the data and the curves calculated with 
these parameters is shown in Fig.~\ref{fig:dsigdt-pp}.
Such results indicate that the PB model is able to describe the $pp$ data, including the dip region. 
However, some comments are in order. The first one concerns the fit to 2.76 TeV data. 
As can be seen in Fig.~\ref{fig:dsigdt-pp} (right panel),
the data cover the range of 0 to 0.7 GeV$^2$, ending in the maximum that follows the dip.
The absence of data at larger values of $|t|$
do not allow a satisfactory determination of $C$ and $D$ parameters in the fit with all free parameters.
To overcome this aspect, we considered a fit to data with $|t|> 0.67$~GeV$^2$ using the parametrization
$\pi C\mathrm{exp}(Dt)$. The result was then used as initial value for the $C$ and $D$ 
parameters in the fit with Eq.~\eqref{eq:pb_modified}. 
We found that the fit with $C$ fixed implies the best description of the data,
showing the expected behaviour of the PB equation at higher values of $|t|$.
The second comment is about the fit of the 13 TeV data. The quite large value of the reduced $\chi^2$ 
obtained in this fit has its origin in the data with $|t| > 2$~GeV$^2$. 
In the right panel of Fig.~\ref{fig:dsigdt-pp}
we see that the distance between the curve and the data points grows with the increasing of $|t|$.
Since the statistical uncertainties of TOTEM data are small, 
these data still contribute significantly to $\chi^2$,
resulting in the value presented in Table~\ref{tab:res-dsigdt-LHC}. 
It is important to emphasize that we have explored the possibility of including the systematic uncertainty in the fit 
(adding it in quadrature to the statistical uncertainty). 
The impact on the parameters is small, but the reduced $\chi^2$ in this case decreases to 2.22, with the description of the data being equivalent to that presented in Fig.~\ref{fig:dsigdt-pp} and the conclusions, discussed below, are not modified.
Finally, we would like to highlight that the PB parameters are mainly determined by the  data 
in the diffraction peak and in the dip-bump region, and such regions of the 13 TeV data are very well described by the model.

\begin{figure}[t]
 \centering
 \includegraphics[scale=0.4]{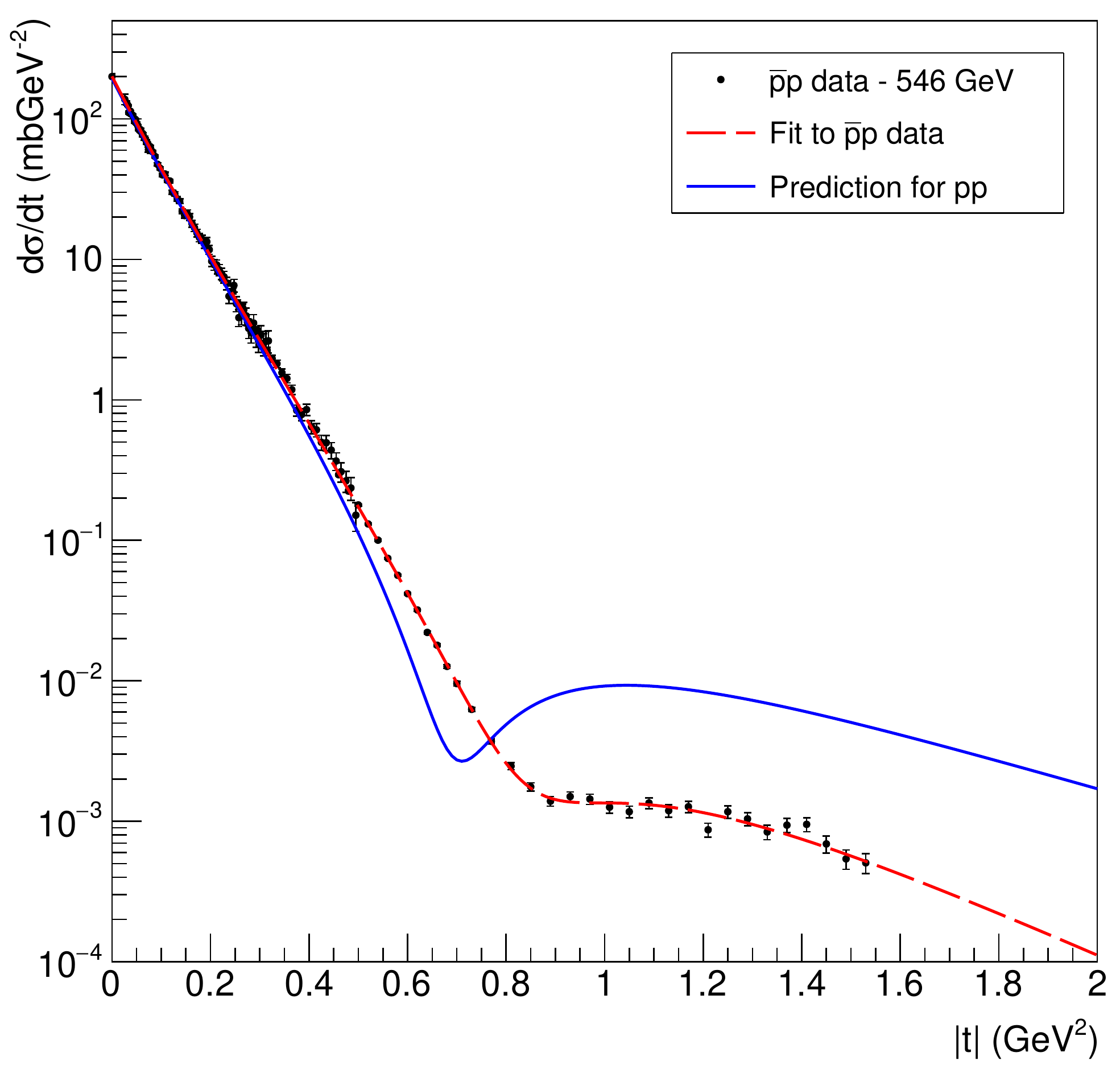}
 \includegraphics[scale=0.4]{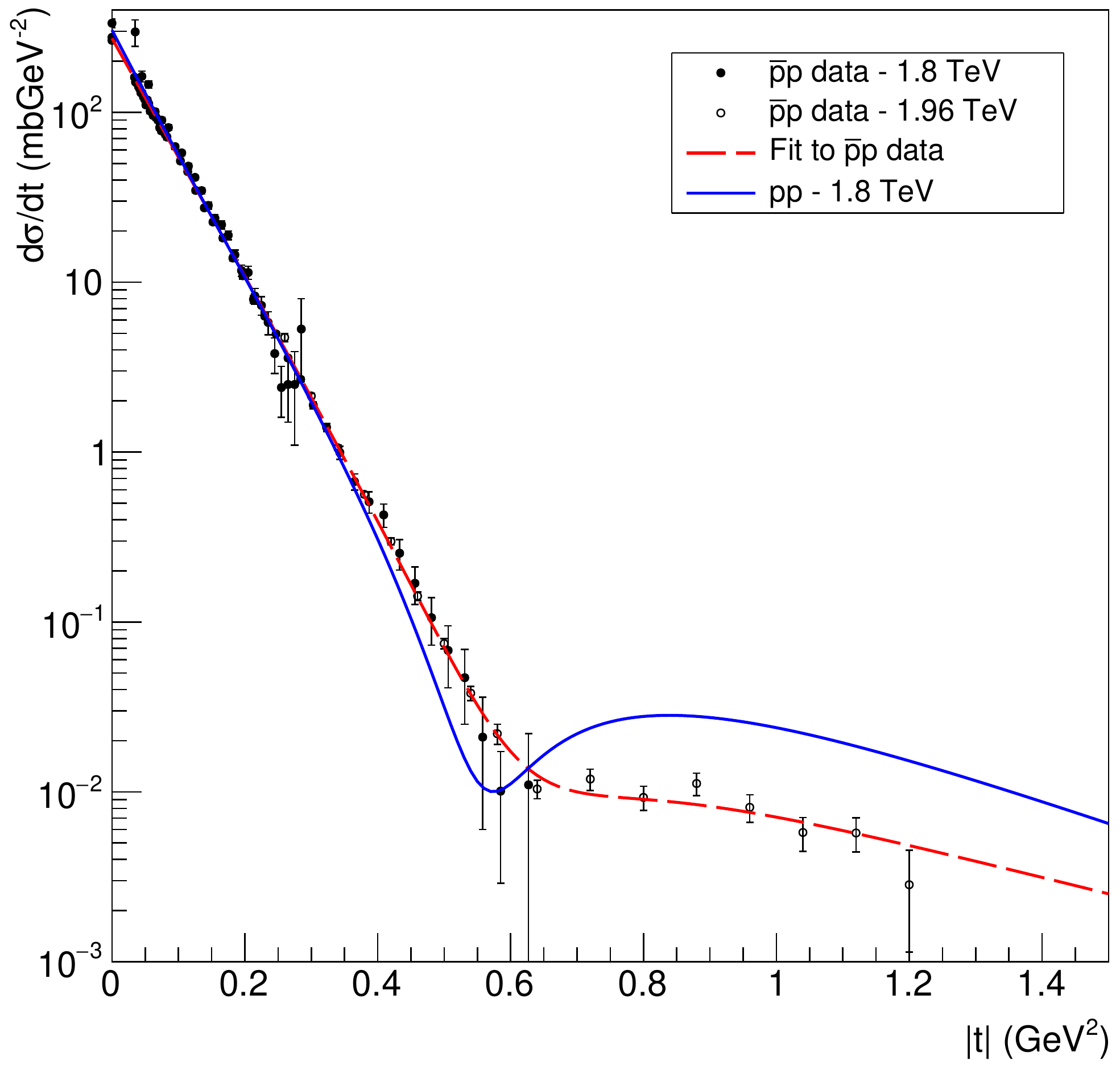}
 \caption{\label{fig:dsigdt_pp_tevatron} Comparison between the PB predictions for the  differential 
cross section for $pp$ collisions at $\sqrt{s} = $ 546 GeV (left) and 1.8 TeV (right) and the experimental  $\bar{p}p$ data. The independent fits of the $\bar{p}p$ data is also presented for comparison.}
\end{figure}

Using the values of the parameters for the different energies presented
in the Tables~\ref{tab:res-dsigdt-ISR} and \ref{tab:res-dsigdt-LHC} 
we will fit $\sqrt{A}$, $B$, $\sqrt{C}$ and $D$ assuming that they 
can be expressed by a function of type: $p_0 + p_1\ln s + p_2 \ln^2 s$. 
On the other hand,  for the parameters $\phi$ and $t_0$ we assume two empirical functions
that become constants at large energies \cite{Fagundes2013}. 
We obtain that:
 \begin{eqnarray}
  \sqrt{A}(s) & = & 3.882 - 0.325\ln s +  0.0331\ln^2 s, \label{eq:par-sqrtA}\\
 B(s) & = & 3.606 + 0.01\ln^2 s,\label{eq:par-B}\\
 \sqrt{C}(s) & = & 0.361 - 0.0884\ln s + 0.00572 \ln^2 s,\label{eq:par-sqrtC}\\
 D(s) & = & 4.730 - 0.642\ln s + 0.036\ln^2 s,\label{eq:par-D}\\
 \phi(s) & = & \dfrac{3.42}{1+\exp(-0.541 \ln s)},\label{eq:par-phi}\\
 t_0(s) & = & \frac{1.026}{1+0.00176\ln^2 s}. \label{eq:par-t0}
 \end{eqnarray}
 We note that, in contrast with the values presented in the Tables~\ref{tab:res-dsigdt-ISR} and \ref{tab:res-dsigdt-LHC}, $\sqrt{A}$ and $\sqrt{C}$ are given in mb 
 in the above equations.  The Eqs.~\eqref{eq:par-sqrtA}-\eqref{eq:par-t0} allow us to calculate $d\sigma/dt$ for the  $pp$ collisions at other center - of - mass energies.
In particular,  we can estimate the PB predictions for the  
  energies of 546 GeV and  1.8 TeV and compare with the existing experimental data for $\bar{p}p$  collisions 
at these energies. This comparison is performed in Fig.~\ref{fig:dsigdt_pp_tevatron}, where  we also present the separate fits of the $\bar{p}p$ data for $\sqrt{s} =$  
546 GeV~\cite{Bozzo1984,Bozzo1985,Abe1994}, 
1.8 TeV~\cite{Amos1990,Abe1994} and 1.96 TeV~\cite{Abazov2012}. 
As in previous studies, the 1.8 TeV and 1.96 TeV data were considered
as being of the same energy (1.8 TeV). The corresponding parameters are presented in the Table~\ref{tab:res-dsigdt-PPBAR}. We have that the PB predictions are not able to describe the $\bar{p}p$ data, with the corresponding values of the reduced $\chi^2/\nu$ being 559.4 and 14.94 for $\sqrt{s} = 546$ GeV and 1.8~TeV, respectively. In contrast, if the data are separately fitted using the PB model, the description is very good, as verified in the Table~\ref{tab:res-dsigdt-PPBAR}. 
Moreover, we have that the PB prediction for the differential cross section presents a dip at $|t| = 0.71$~GeV$^2$ for $\sqrt{s}$ = 546 GeV 
and $|t| = 0.57$~GeV$^2$ for 1.8 TeV, followed by a bump,  in contrast with the $\bar{p}p$ data.
The large difference between the PB predictions, constrained by $pp$ data, and the experimental $\bar{p}p$ data implies that the scattering amplitude for the $pp$ and $\bar{p}p$ processes is different for a given energy, which can be interpreted as being due to the Odderon contribution. Therefore, our results suggest the presence of the Odderon in the description of the hadronic collisions at high energies.  
We stress that in order to properly estimate the Odderon contribution, it is fundamental to have data  of both 
$pp$ and $\bar{p}p$ scattering at the same energy.
Our analysis only provides more one indication of its presence at high energies.

\begin{table}[t]
 \centering
  \begin{tabular}{|c||c|c|}\hline\hline
 \multicolumn{3}{|c|}{$\bar{p}p$ scattering}\\\hline
 Energy                   & 546 GeV    & 1.8 TeV + 1.96 TeV  \\\hline
 $A$                  & 67.51(29)  & 88.5(1.3) \\
 $B$                   & 5.103(23)  & 13.0(1.9) \\
 $C (\times 10^{-3})$   & 59.4(7.2)  & 24(14)    \\
 $D$                    & 3.67(10)   & 2.26(58)  \\
 $\phi$                 & 3.6087(94) & 4.08(12)  \\
 $t_0$                  & 0.7820(80) & 2.7(1.8)  \\\hline
 $\chi^2/\nu$             & 1.097      & 2.365     \\
 $\nu$                       & 130        & 91        \\\hline\hline
 \end{tabular} 
\caption{Results for the fits of the  differential 
 cross section data considering $\bar{p}p$ collisions. The reduced $\chi^2$ and degrees of freedom, $\nu$, for each fit are also presented.}
\label{tab:res-dsigdt-PPBAR}
 \end{table}

As the energy dependence of the parameters present in PB model have been determined, we also can estimate the total $pp$ cross sections for other values of the center - of - mass energy. In Table~\ref{tab:pred_sigtot} we present our predictions for 
$\sigmatot^{pp}$ for the values of $\sqrt{s}$ in which experimental $\bar{p}p$ data are available. 
The uncertainties quoted in our predictions have been  obtained by means
of error propagation from the uncertainties and correlation
coefficients of the parameters present in Eqs.~\eqref{eq:par-sqrtA}-\eqref{eq:par-t0}.
We note that our prediction for 546 GeV is compatible with
the experimental  $\bar{p}p$ data \cite{Abe1994b}.
On the other hand, at 900 GeV, in which new measurements are expected at the LHC,
we find a cross section that is larger than the measured in $\bar{p}p$ collisions at  the same energy \cite{Alner1986}. It is important to emphasize that the PB model predicts that  the differential cross section  for $pp$ collisions at $\sqrt{s} = $ 900 GeV will have a dip in transferred momentum around 0.64 GeV$^2$. Our prediction for  $\sqrt{s} = $ 1.8 TeV is above the experimental 
 $\bar{p}p$ values obtained by the E-710 and E-811 Collaborations~\cite{Amos1992,Avila2002} and is slightly smaller than the  CDF data~\cite{Abe1994b}.
These results point to $\sigmatot^{pp} > \sigmatot^{\bar{p}p}$ for energies
greater than 546 GeV, a scenario possible to occur once the Odderon contribution exists.
In this case, the difference between these cross sections is bounded,
$|\sigmatot^{pp}-\sigmatot^{\bar{p}p}|< \ln s$.
This result also indicates that the imaginary part of the Odderon amplitude in the forward direction 
is positive at high energies.  
We have also calculated $\sigmatot^{pp}$ at 14 TeV and obtained $113.3 \pm 0.1$~mb, with the differential cross section at this energy presenting a dip for a squared transferred momentum close to 0.45 GeV$^2$.
%

\begin{table}[t]
 \centering
 \begin{tabular}{|c||c|c|}\hline\hline
  Energy  & $\sigmatot$ ($pp$) & Experimental value ($\bar{p}p$)\\\hline
  546 GeV & 61.6 $\pm$ 0.2 mb  & 61.26 $\pm$ 0.93 mb (CDF \cite{Abe1994b})  \\\hline
  900 GeV & 67.6 $\pm$ 0.2 mb  & 65.3 $\pm$ 1.6 mb (UA5 \cite{Alner1986}) \\\hline
          &                    & 72.8 $\pm$ 3.1 mb (E-710 \cite{Amos1992})\\
  1.8 TeV & 77.2 $\pm$ 0.2 mb  & 71.42 $\pm$ 2.42 mb (E-811 \cite{Avila2002}) \\
          &                    & 80.03 $\pm$ 2.24 mb (CDF \cite{Abe1994b})  \\\hline\hline
 \end{tabular}
 \caption{\label{tab:pred_sigtot}Predictions of the PB model for the total cross section considering  $pp$ collisions at different values of the center - of - mass energy. For comparison, the experimental results for $\bar{p}p$ scattering are also presented.}
\end{table}

In summary, we have analyzed the experimental data for the differential cross section
in elastic  $pp$ collisions using the model proposed by  Phillips and Barger in the 1970's and recently improved by the inclusion of the proton form factor in order to describe data in the low $|t|$ region. We have demonstrated that the PB model is able to describe the recent LHC data and  derived the energy dependence of its parameters. Predictions for $d\sigma/dt$ and $\sigmatot$ considering $pp$ collisions at different values of $\sqrt{s}$ have been 
presented and a comparison with the available $\bar{p}p$ data was performed. We have verified that the PB model implies 
 $\sigmatot^{pp} > \sigmatot^{\bar{p}p}$
 and the presence of a dip in $pp$ collisions at 
 $\sqrt{s} = 546$ GeV and 1.8~TeV, which is not present in the $\bar{p}p$ data. These results indicate that, for a fixed energy, the scattering amplitudes for the two crossed channels are different, which can be interpreted as being due to the Odderon contribution. Therefore, our result can be considered more one indication that the Odderon should be considered in the description of the hadronic collisions at high energies. The future LHC data at $\sqrt{s} = 900$ GeV probably will allows to establish a definitive conclusion.

\section*{Acknowledgements}
VPG acknowledge very useful discussions  about elastic scattering with Marcio Menon, Daniel Fagundes and Anderson Kendi Kohara along the last years. This work was  partially financed by the Brazilian funding
agencies CNPq,  FAPERGS and INCT-FNA (process number 
464898/2014-5).




\begin{thebibliography}{99}


\bibitem{Antchev2011} G. Antchev \textit{et al.} (TOTEM Collab.), EPL \textbf{95}, 41001 (2011).
\bibitem{Antchev2013} G. Antchev \textit{et al.} (TOTEM Collab.), EPL \textbf{101}, 21002 (2013).

\bibitem{Antchev2015} G. Antchev \textit{et al.} (TOTEM Collab.), Nucl. Phys. B \textbf{899}, 521 (2015).
\bibitem{Antchev2016} G. Antchev \textit{et al.} (TOTEM Collab.), Eur. Phys. J. C \textbf{76}, 661 (2016).
\bibitem{Deile2017} M. Deile, Talk at WE-Heraeus Physics School: QCD - Old Challenged and New Opportunities (Sept 24-30, 2017);\\ \url{https://indico.cern.ch/event/614845/contributions/2728919/}.



\bibitem{Antchev2017b} G. Antchev \textit{et al.} (TOTEM Collab.), Eur. Phys. J. C \textbf{79}, 103 (2019).

 \bibitem{Antchev2017} G. Antchev \textit{et al.} (TOTEM Collab.), CERN-EP-2017-335.

 
\bibitem{Antchev2018a} G. Antchev \textit{et al.} (TOTEM Collab.), CERN-EP-2018-338; arXiv:1812.08283 [hep-ex].

\bibitem{Antchev2018b} G. Antchev \textit{et al.} (TOTEM Collab.), CERN-EP-2018-341; arXiv:1812.08610 [hep-ex].
 
 
\bibitem{Aad:2014} 
  G.~Aad \textit{et al.} (ATLAS Collab.),
  Nucl.\ Phys.\ B \textbf{889}, 486 (2014);
  arXiv:1408.5778 [hep-ex].

\bibitem{Aaboud:2016} 
  M.~Aaboud \textit{et al.} (ATLAS Collab.),
  Phys.\ Lett.\ B \textbf{761}, 158 (2016); 
  arXiv:1607.06605 [hep-ex].
 


\bibitem{Pancheri:2016yel} 
  G.~Pancheri and Y.~N.~Srivastava,
  Eur.\ Phys.\ J.\ C {\bf 77}, no. 3, 150 (2017)


\bibitem{Martynov:2017zjz} 
  E.~Martynov and B.~Nicolescu,
  Phys.\ Lett.\ B {\bf 778}, 414 (2018)

\bibitem{Khoze:2018bus} 
  V.~A.~Khoze, A.~D.~Martin and M.~G.~Ryskin,
  Phys.\ Lett.\ B {\bf 780}, 352 (2018)
  
  \bibitem{Broilo:2018els} 
  M.~Broilo, E.~G.~S.~Luna and M.~J.~Menon,
  Phys.\ Lett.\ B {\bf 781}, 616 (2018)
  
  \bibitem{Troshin:2018ihb} 
  S.~M.~Troshin and N.~E.~Tyurin,
  Mod.\ Phys.\ Lett.\ A {\bf 33}, 1850206 (2018)
  
  
  \bibitem{Broniowski:2018xbg} 
  W.~Broniowski, L.~Jenkovszky, E.~Ruiz Arriola and I.~Szanyi,
  Phys.\ Rev.\ D {\bf 98}, no. 7, 074012 (2018)
  
  \bibitem{Broilo:2018qqs} 
  M.~Broilo, E.~G.~S.~Luna and M.~J.~Menon,
  Phys.\ Rev.\ D {\bf 98}, no. 7, 074006 (2018)


\bibitem{Gotsman:2018buo} 
  E.~Gotsman, E.~Levin and I.~Potashnikova,
  Phys.\ Lett.\ B {\bf 786}, 472 (2018)
  
\bibitem{Csorgo:2018uyp} 
  T.~Csorgo, R.~Pasechnik and A.~Ster,
  Eur.\ Phys.\ J.\ C {\bf 79}, no. 1, 62 (2019); 
  arXiv:1811.08913 [hep-ph].


 \bibitem{Ewerz} C.~Ewerz,
 arXiv:hep-ph/0306137.

\bibitem{dosch_ewerz} 
  H.~G.~Dosch, C.~Ewerz and V.~Schatz,
  Eur.\ Phys.\ J.\ C {\bf 24}, 561 (2002)

\bibitem{Ster:2015esa} 
  A.~Ster, L.~Jenkovszky and T.~Csorgo,
  Phys.\ Rev.\ D {\bf 91}, no. 7, 074018 (2015)




%


 
 
 
 
 \bibitem{Phillips1973} R.J.N. Phillips, V.D. Barger, Phys. Lett. \textbf{46B}, 412 (1973).
 
 \bibitem{Fagundes2013} D.A. Fagundes, G. Pancheri, A. Grau, S. Pacetti, Y.N. Srivastava, Phys. Rev. D \textbf{88}, 094019 (2013).
 
 \bibitem{AmaldiSchubert1980} U. Amaldi, K.R. Schubert, Nucl. Phys. B \textbf{166}, 301 (1980).
 
 \bibitem{Schubert1979} K.R. Schubert, \textit{Landolt - B\"ornstein, Numerical Data and Functional Relationships in Science and Technology, New Series}, Vol. I/9a (Springer-Verlag, Berlin, 1979).



 \bibitem{Bozzo1984} M. Bozzo \textit{et al.} (UA4 Collab.), Phys. Lett. B \textbf{147}, 385 (1984).
 \bibitem{Bozzo1985} M. Bozzo \textit{et al.} (UA4 Collab.), Phys. Lett. B \textbf{155}, 197 (1985).
 
 \bibitem{Abe1994} F. Abe \textit{et al.} (CDF Collab.), Phys. Rev. D \textbf{50}, 5518 (1994). 

 
 \bibitem{Amos1990} N.A. Amos \textit{et al.} (E-710 Collab.), Phys. Lett. B \textbf{247}, 127 (1990).
 
 \bibitem{Abazov2012} V.M. Abazov \textit{et al.} (D0 Collab.), Phys. Rev. D \textbf{86}, 012009 (2012).
%


\bibitem{Abe1994b} F.~Abe \textit{et al.} (CDF Collab.), Phys.\ Rev.\ D \textbf{50}, 5550 (1994).


\bibitem{Alner1986} G.J. Alner \textit{et al.} (UA5 Collab.) Z. Phys. C \textbf{32}, 153 (1986).



\bibitem{Amos1992} N.~A.~Amos \textit{et al.} (E-710 Collab.), Phys.\ Rev.\ Lett.\  \textbf{68}, 2433 (1992).
  
\bibitem{Avila2002} C.~Avila \textit{et al.} (E-811 Collab.), Phys.\ Lett.\ B \textbf{537}, 41 (2002).
  



\end{thebibliography}
\end{document}